\documentclass[11pt]{article}
\usepackage{moriond,epsfig}

\def\Journal#1#2#3#4{{#1} {\bf #2}, #3 (#4)}

\def\PRL{\em Phys. Rev. Lett.}

\def\JHEP{{\em JHEP}}

\def\be{\begin{equation}}
\def\ee{\end{equation}}
\def\bea{\begin{eqnarray}}
\def\eea{\end{eqnarray}}

\begin{document}
\vspace*{3cm}
\title{The ATLAS discovery reach for SUSY models with early data}

\author{Janet Dietrich (for the ATLAS collaboration)}

\address{Physikalisches Institut, Albert-Ludwigs-Universit\"at Freiburg, Germany\\
        E-mail: janet.dietrich@cern.ch}

\maketitle\abstracts{The search for physics beyond the Standard Model (BSM) is one 
of the most important goals for the general purpose detector ATLAS at the Large Hadron Collider at CERN. We review some of the current strategies to search for generic SUSY models with R-parity conservation in channels with jets, leptons and missing transverse energy for an integrated luminosity of ${\cal L}$ = 200~${pb}^{-1}$ at a centre-of-mass energy $\sqrt{s}$ = $10$~TeV. Only a selection of the results is presented with a focus on the discovery potential for inclusive searches. 
The discovery reach for a centre-of-mass energy of $\sqrt{s}$ = $7$~TeV and an integrated luminosity of ${\cal L}$ = 1$~{fb}^{-1}$ is expected to be similar to the one discussed in this note.}
\section{Inclusive searches for SUSY signals}
Searches for supersymmetry (SUSY) have to deal with models with a large set of free parameters. 
In this article we consider only the R-parity conserving scenario in which the lightest SUSY particle (LSP) is stable. Here, the LSP is a neutral particle that is produced at the end of the cascade decays of massive supersymmetric particles and escapes detection, causing large missing transverse energy $E_{T}^{miss}$. Thus a requirement of large $E_{T}^{miss}$ in the event along with other final state particles such as leptons and/or jets with high $p_{T}$ has a high potential to discover SUSY at the LHC.
\\
The primary goal of the presented work was to study different decay topologies and signatures for many ``typical'' MSSM points, i.e. a large range of possible mass patterns. About 200 MSSM points were selected to scan the MSSM parameter space.
Since there is no unique model of SUSY-breaking, all these models should be viewed
only as possible patterns of LHC signatures, not as complete theories.
\paragraph{mSUGRA models}
In order to cover a large parameter space and to reduce the number of SUSY points minimal SUper GRAvity (mSUGRA~\cite{msugra}) grids were made in "radial coordinates", i.e. points on outgoing radial lines in the $(m_0,m_{1/2})$ plane for $\tan \beta =$ 10. Lines with different slopes were produced. The other mSUGRA parameters are $A_0=0$, $\mu >0$.
\paragraph{pMSSM models}
About 150 points were generated in the phenomenological MSSM (pMSSM~\cite{pMSSM}) space with 19 free soft SUSY breaking parameters. This parameter space was sampled with a flat prior distribution within certain theoretical limits and a mass scale of $< 1$ TeV. Only points are chosen which satisfied various experimental bounds like constraints from collider experiments at LEP and Tevatron, but also the WMAP dark matter density upper bound and bounds from direct DM detection searches~\cite{tom_rizzo}.\\

ATLAS studied various search channels with different numbers of jets ($\ge 2$, $\ge 3$, $\ge 4$) and leptons (0, 1, 2), trying to keep the SUSY searches robust in order to discriminate a potential SUSY signal from the main background sources. The leptons considered are either a muon or an electron; the number of leptons defines mutually exclusive channels:

\paragraph{0-lepton channels}
In the no-lepton search mode events with an isolated high $p_{T}$ ($>20$~GeV) electron
or muon are vetoed. The dominant backgrounds in the 0 lepton channel are the $QCD$ multijets, $Z \to \nu\bar{\nu}$+jets, $W$+jets and $t\bar{t}$ production. In general, this channel shows the highest discovery potential, but requires that the $QCD$ background and sources of fake $E_{T}^{miss}$ can be kept under control. 
\paragraph{1-lepton channels}
For the 1-lepton mode one identified high $p_T$ lepton is required and events with 2 leptons are vetoed. In this channel the $QCD$ and $Z$+jets background are strongly reduced by the lepton requirement. The main backgrounds are $t\bar{t}$ decays and $W$+jets production. 
\paragraph{2-lepton channels}
The 2-lepton channel is subdivided into two channels: one asking for opposite-sign electric charges (non-negligible SM background, but high SUSY signal statistics) and one asking for same-sign electric charges (low SM background, but lower SUSY signal statistics). The main possible SM processes that can mimic these final states are the production of $t\bar{t}$, $W$+jets, $Z$+jets and Diboson processes. \\

Further interesting search modes in ATLAS that are not presented in this note include channels containing either photons, tau leptons, b-jets or $\geq$ 3 leptons~\cite{ATLAS}'~\cite{gmsb_note}'~\cite{b_note}.\\
All cuts on the number of jets and on the transverse momentum of jets are common to the channels with different lepton multiplicities and are summarized in Table~\ref{table_njets_definition}. 
The $p_T$ cuts on the jets are chosen in order to be consistent with the multijet trigger requirements and in order to reject a sufficient amount of background.
\begin{table}[h]
  \begin{center}
    \begin{tabular}{|l|c|c|c|}
      \hline
      Number of jets                           & $\ge 2$ jets    & $\ge 3$ jets        & $\ge 4$ jets        \\ 
      \hline
      Leading jet $p_{T}$ (GeV)    & $>180$       & $>100$           & $>100$           \\
      Jets $p_{T} $ (GeV) & $>50$ (Jet 2)        &  $>40$ (Jet 2-3)            & $>40$ (Jet 2-4)            \\
     $\vert \Delta\phi(jet_{i},E_{T}^{miss})\vert$       & [$>0.2$,$>0.2$] & [$>0.2$,$>0.2$,$>0.2$] & [$>0.2$,$>0.2$,$>0.2$,$>0.0$] \\ 
     $E_{T}^{miss} > f\times M_{eff}$         & $f=0.3$       & $f=0.25$          & $f=0.2$           \\       
    \hline
   \end{tabular}
   \caption{The table shows the cuts for the 3 studied jet multiplicities: cuts on the $p_{T}$ of the leading, the $p_{T}$ of the remaining jets, $\vert\Delta\phi(jet_{i},E_{T}^{miss})\vert$, $M_{eff}$ fraction $f$}
    \label{table_njets_definition}
  \end{center}
\end{table}

\begin{figure*}[htb]
\centering
\resizebox{67mm}{62mm}{\includegraphics{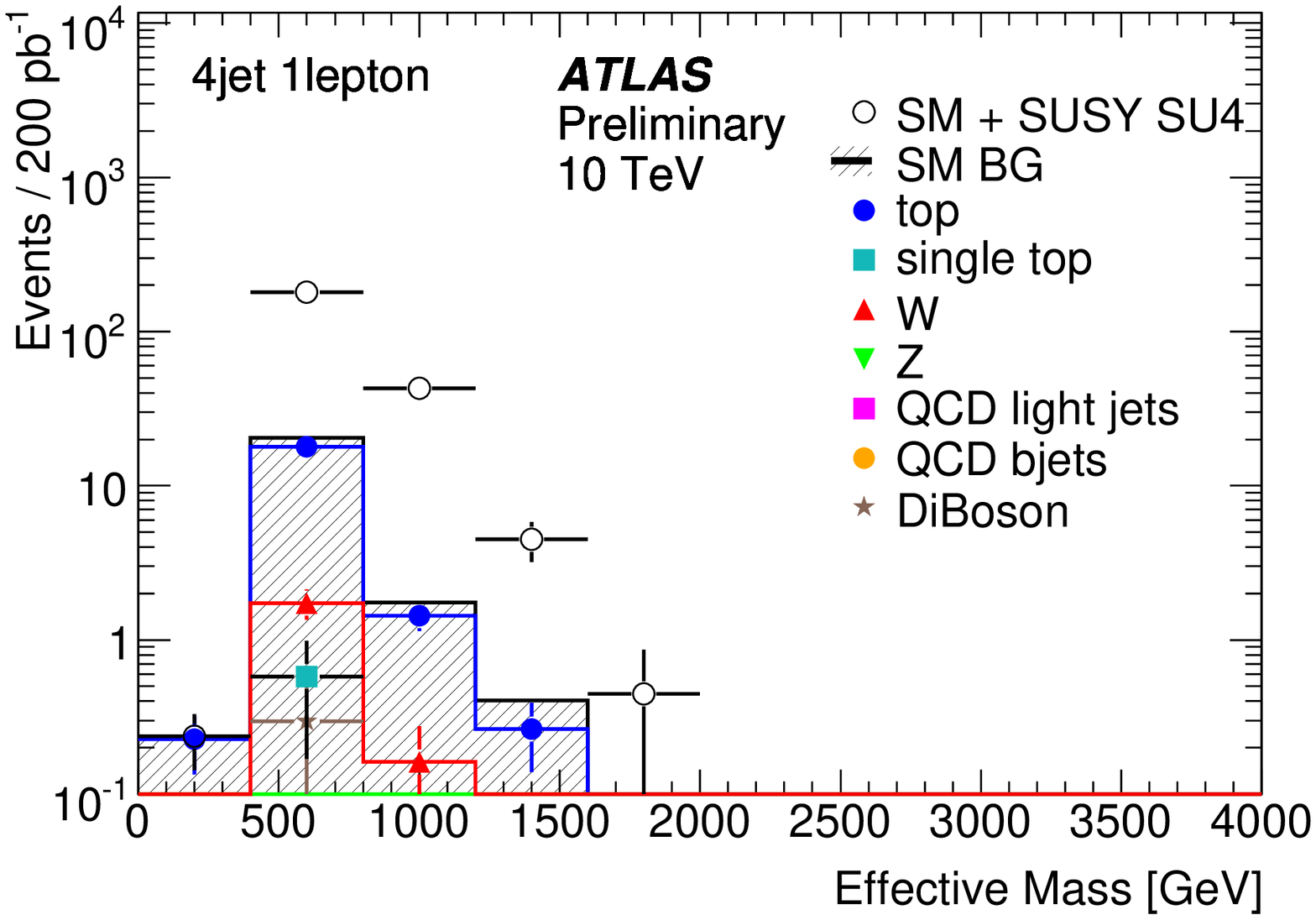}}
\resizebox{69mm}{62mm}{\includegraphics{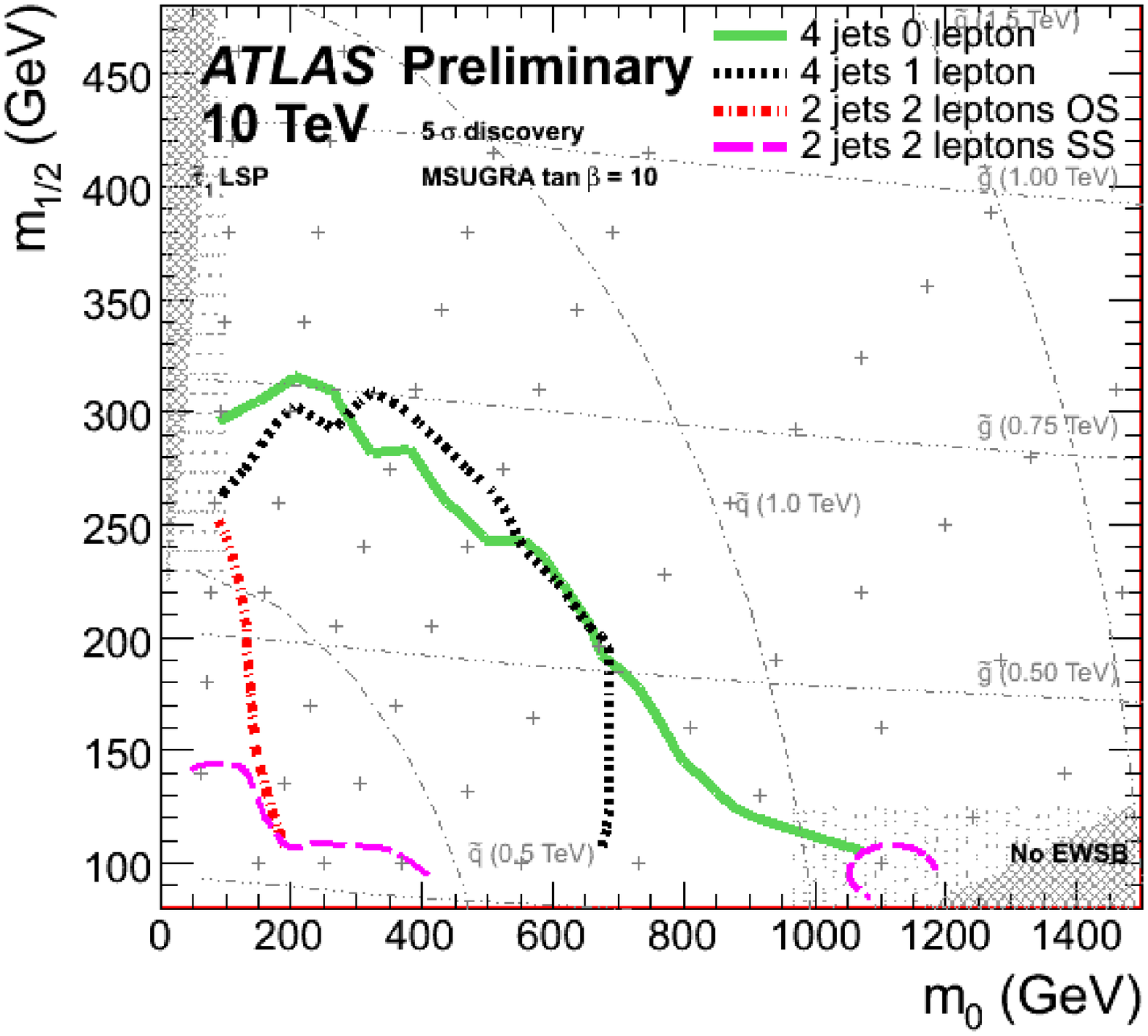}}

\caption{The effective mass $M_{\rm{eff}}$ distribution after the final selection for the inclusive 4-jet 1 lepton channel (left side). The right plots shows the 5$\sigma$ contour lines for the ATLAS experiment for 200~${pb}^{-1}$ at $\sqrt{s} =10$~TeV for the 4-jet 0 lepton, the 4-jet 1 lepton and the 2-jet 2 lepton channel as a function of $M_0$ and $M_{1/2}$ for the mSUGRA model with $\tan \beta =10$.} 
\label{fig_meff_1lep}
\end{figure*}

In addition to the cuts described in Table~\ref{table_njets_definition} the following cuts are applied: $E_{T}^{miss} > 80$~GeV and transverse sphericity $S_{T}$~\footnote{$S_{T}$ is defined as: $$S_T\equiv\frac{2\lambda_2}{(\lambda_1+\lambda_2)}$$ where $\lambda_1$ and $\lambda_2$ are the eigenvalues of the $2\times2$ sphericity tensor \mbox{$S_{ij}=\sum_kp_{ki}p^{kj}$} computed from all selected jets and leptons~\cite{ATLAS}.}
$> 0.2$. Furthermore, for the one lepton channel the requirement of the transverse mass $M_{T}$, constructed from the identified lepton and the missing transverse energy, must be greater than 100~GeV. A full description of all selection cuts is presented in the ATLAS note~\cite{atlas_pub_note}'~\cite{wisconsin}.
The final variable we have used to discriminate between SUSY and background is the effective mass ($M_{\rm{eff}}$) defined as the scalar sum of the transverse momenta of the main objects in the event: $M_{eff} \equiv \sum_{i=1}^{N_{jets}} P_{T}^{{\rm jet},i} + \sum_{i=1}^{N_{lep}} P_{T}^{{\rm lep},i} + E_{T}^{miss}$
where $N_{jets}$ is the number of jets (2 - 4) and $N_{lep}$ is the number of leptons (0 - 2). Further high $p_T$ jets or leptons are not included in the sum. 
Figure \ref{fig_meff_1lep} shows the $M_{eff}$ distribution for the 4-jet 1 lepton channel after applying all cuts for the ATLAS benchmark point SU4~\footnote{$m_0=200$~GeV, $m_{1/2}=160$~GeV, $A_0=-400$~GeV, $\tan \beta = 10$, $\mu >0$; $\sigma$ = $108$~pb at $\sqrt{s} =10$~TeV.}; the signal clearly exceeds the background in both channels.

\section*{Discovery reach}
We have explored the reach of our search strategies by studying grids of models in the parameter space of mSUGRA and pMSSM. For each point in the two grids the same set of selection cuts~\footnote{The $M_{eff}$ cut was varried for every point with a step size of 400~GeV, the cut that gives the highest significance $Z_{N}$ was applied.} are applied and the significance $Z_{N}$ is calculated. This $Z_N$ method uses a convolution of a Poisson and a Gaussian term to account for the systematic uncertainty~\cite{ATLAS}. 
Note that for the studies presented here, a systematic uncertainty of $50\%$ was assigned to the SM background estimate.
The following plots show only the channels with the largest discovery reach for each lepton 
multiplicity. 
No attempt was made to combine the significance of the various channels. 
The 5$\sigma$ discovery reach lines in the $M_{0}$ - $M_{1/2}$ plane for the 4-jet 0 lepton, 4-jet 1 lepton and 2-jet 2 lepton channel for the mSUGRA model are presented in Figure~\ref{fig_meff_1lep} in the right plot. From this Figure, one can see that the 0 and 1 lepton channels have similar potential for a discovery in the studied mSUGRA grids. 
\begin{figure*}[htb]
\centering
\includegraphics[height=78mm]{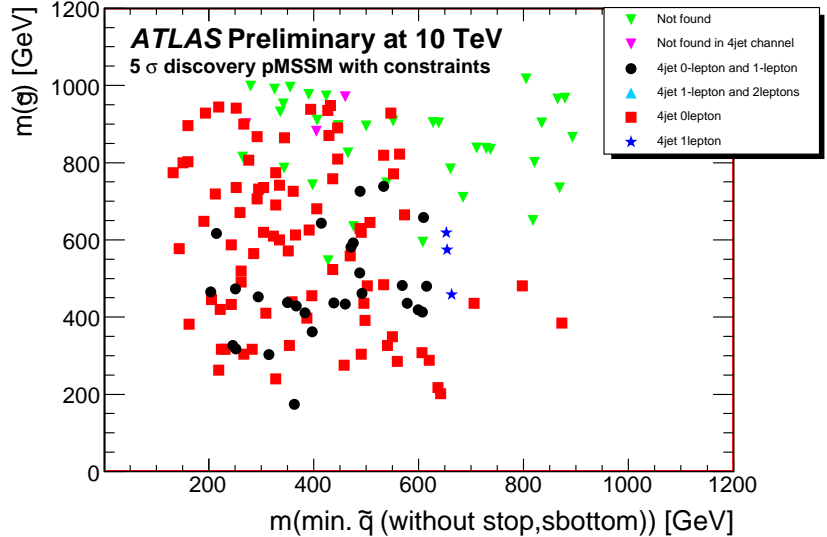}
\caption{The 5$\sigma$ contour lines for the ATLAS experiment for 200~${pb}^{-1}$ at $\sqrt{s} =10$~TeV for the 4-jet 0 lepton, the 4-jet 1 lepton and the 2 lepton channels for the points of the pMSSM grids with constraints as a function of the minimal mass of the light squarks and the gluino mass.} 
\label{reach_msugra}
\end{figure*}
Figure~\ref{reach_msugra} shows the 5$\sigma$ discovery reach for the pMSSM grid with constraints as a function of the minimum mass of the first and second generation squarks and the mass of the gluino. Most considered SUSY signals can be discovered with the 4-jet channels if the cross section is larger than $10$~pb and if the squark or gluino mass are $\leq 600-700$~GeV.
A few points are only found with the 2- or 3-jet channels with 0 or 1 lepton.
In general the 4-jet 0-lepton channel is more effective than the 1-lepton one, because many points do not lead to significant high $p_T$ lepton production. 
\section*{Conclusion}
The discovery potential for inclusive SUSY search channels with 0 lepton, 1 lepton or 2 leptons and $\ge 2$, $\ge 3$ or $\ge 4$ jets has been investigated for a scenario assuming an LHC centre-of-mass energy of 
$\sqrt{s}$ = $10$~TeV and an integrated luminosity of ${\cal L}$ = 200~${pb}^{-1}$.
The results of the scans presented indicate that ATLAS could discover
signals of R-parity conserving SUSY with gluino and squark masses less than 600 -700~GeV in many scenarios. 

\section*{Acknowledgments}
The author would like to thank the organisers of the XLVth Rencontres de Moriond session for the invitation to present her work and everyone in the ATLAS Collaboration whose work contributed to this talk. The author acknowledges the support by the Landesstiftung Baden W\"urttemberg and the BMBF.

\end{document}